\documentclass[10pt, conference, letterpaper]{IEEEtran}
\IEEEoverridecommandlockouts

\usepackage{caption}
\usepackage{subcaption}
\usepackage{url}
\usepackage{cite}
\usepackage{amsmath,amssymb,amsfonts}
\usepackage{algorithmic}
\usepackage{graphicx}
\usepackage{textcomp}
\usepackage{xcolor}
\usepackage{tabto}
\usepackage{comment}
\usepackage{booktabs}
\usepackage{diagbox}
\usepackage{flexisym}
\usepackage{authblk}

\usepackage{tikz}
\usepackage{eso-pic}

\DeclareMathOperator{\E}{\mathbb{E}}
\def\BibTeX{{\rm B\kern-.05em{\sc i\kern-.025em b}\kern-.08em
    T\kern-.1667em\lower.7ex\hbox{E}\kern-.125emX}}
\begin{document}

%

\newcommand\copyrighttext{%
IEEE Copyright Notice
\vspace{0.05in}

  \footnotesize \textcopyright 2021 IEEE. Personal use of this material is permitted.
  Permission from IEEE must be obtained for all other uses, in any current or future
  media, including reprinting/republishing this material for advertising or promotional
  purposes, creating new collective works, for resale or redistribution to servers or
  lists, or reuse of any copyrighted component of this work in other works.}
\newcommand\copyrightnotice{%
\begin{tikzpicture}[remember picture,overlay]
\node[anchor=south,yshift=10pt] at (current page.south) {\fbox{\parbox{\dimexpr\textwidth-\fboxsep-\fboxrule\relax}{\copyrighttext}}};
\end{tikzpicture}%
}

\newcommand\AtPageUpperMyright[1]{\AtPageUpperLeft{
 \put(\LenToUnit{0.099\paperwidth},\LenToUnit{-1cm}){
     \parbox{0.9\textwidth}{\raggedleft\fontsize{9}{11}\selectfont #1}}
 }}
\newcommand{\conf}[1]{
\AddToShipoutPictureBG*{
\AtPageUpperMyright{#1}
}
}

\title{Cocktail Edge Caching: Ride Dynamic Trends of Content Popularity with Ensemble Learning\\
}
\conf{Accepted to be Published in: IEEE International Conference on Computer Communications (INFOCOM), May 10$-$13, 2021
\noindent\makebox[\linewidth]{\rule{0.755\paperwidth}{0.5pt}}}
\author{Tongyu Zong$^1$,\enspace Chen Li$^1$,\enspace Yuanyuan Lei$^1$,\enspace Guangyu Li$^1$,\enspace Houwei Cao$^2$,\enspace and Yong Liu$^1$\\$^1$Tandon School of Engineering, New York University, USA\\$^2$Department of Computer Science, New York Institute of Technology, USA}



\maketitle
\copyrightnotice

\vspace{-0.14in}
\begin{abstract}
Edge caching will play a critical role in facilitating the emerging content-rich applications. However, it faces many new challenges, in particular, the highly dynamic content popularity and the heterogeneous caching configurations. In this paper, we propose Cocktail Edge Caching, that tackles the dynamic popularity and heterogeneity through ensemble learning. Instead of trying to find a single dominating caching policy for all the caching scenarios, we employ an ensemble of constituent caching policies and adaptively select the best-performing policy to control the cache. Towards this goal, we first show through formal analysis and experiments that different variations of the LFU and LRU policies have complementary performance in different caching scenarios. We further develop a novel caching algorithm that enhances LFU/LRU with deep recurrent neural network (LSTM) based time-series analysis. Finally, we develop a deep reinforcement learning agent that adaptively combines base caching policies according to their virtual hit ratios on parallel virtual caches. Through extensive experiments driven by real content requests from two large video streaming platforms, we demonstrate that CEC not only consistently outperforms all single policies, but also improves the robustness of them. CEC can be well generalized to different caching scenarios with low computation overheads for deployment.
\end{abstract}

\vspace{-0.1in}
\section{Introduction}
Online services, ranging from web hosting, video streaming, gaming, to Virtual/Augmented/Mixed Reality (VR/AR/MR), etc., are increasingly dependent on the timely delivery of rich media content over the global Internet. 
Content delivery systems are facing new challenges. First of all, the emerging new content requires orders of magnitude higher bandwidth. A  premium quality 360 degree video can easily consume a bandwidth of multiple Gigabits-per-second (Gbps)~\cite{Huawei_Report}. Secondly, many new multimedia applications also involve live interaction between users, which requires low-latency content delivery. Finally, content popularity becomes more and more dynamic. User-Generated Content (UGC) has become tremendously popular on platforms like TikTok, Twitch and YouTube, etc.  A new UGC item may suddenly go viral and attract a flash crowd of viewers to watch it within a short time period.
  
Caching, an {\it ``old trick"} from the early days of the Internet, will continue to play a critical role in the contemporary content delivery ecosystem~\cite{In-network,jacobson2009networking,FemtoCaching,sun2020flocking}. 
By placing popular content at the network edge, {\bf Edge Caching} helps users retrieve content at high throughput and low latency, while reducing the traffic in the core network.  
Each edge cache box necessarily has less resources, in particular smaller buffer, than the traditional central cache server. The traditional caching policies are driven by simple statistics of request history, e.g. the past request frequencies or the time elapsed since the last request. When the number of users served by a cache server is large, those simple statistics serve as  reliable future  popularity predictors to drive simple caching policies, such as Least-Frequently-Used (LFU) or Least-Recently-Used (LRU), thanks to the ``Law-of-Large-Numbers" effect. An edge cache box only serves a small user population, whose aggregate content consumption patterns are much more volatile and highly non-stationary. To complicate the matters further, caching scenarios at different levels of the caching hierarchy are heterogeneous in terms of buffer sizes and user/content mixtures.  Simple static caching policies are inadequate to deliver good edge caching performance. 

In this study, we tackle the dynamic content popularity and configuration heterogeneity problems in edge caching through {\bf ensemble learning}. The general idea of ensemble learning is to strategically generate and combine multiple models to improve the performance of a single model and/or reduce the likelihood of selecting a poor model. It has wide applications in model selection, data fusion, and confidence estimation, as surveyed in  \cite{Sagi2018ensemble}. As a closely-related example, the winner of the famous Netflix Prize~\cite{NetflixPrize} used  ensemble learning to predict content preference of individual users, where several base recommendation models were combined to predict a final single user-content rating that exploits the strengths of each model~\cite{WinningNetflix}. Cache hit ratio hinges on the accuracy of predicting future content popularity among the group of users served by a cache. We use ensemble learning to ride dynamic content popularity trends and improve edge caching hit ratio. 

Edge caching scenarios are complicated due to different content and user mixtures, time-varying popularity evolution, and heterogeneous  cache configurations. {\it Instead of trying to find a single dominating caching policy for all the caching scenarios, we  employ an ensemble of constituent caching policies that concurrently process content requests through {\bf virtual caches}, and adaptively select the best-performing policy to control the primary cache by a Deep Reinforcement Learning (DRL) agent.} The DRL agent is trained to combine the merits of constituent caching policies to address the complicated caching scenarios. The high-level idea is similar to the classic combination therapy, namely AIDS cocktail~\cite{henkel1999attacking}. We call our framework {\bf Cocktail Edge Caching (CEC)}. 
%
Towards developing CEC, we made the following contributions:
\begin{enumerate}
\item We first formulate caching under dynamic content  popularity as a future popularity estimation problem. We then formally study the sliding-window based variations of the classic LFU and LRU policies. Through analysis and experiments, we demonstrate the impact of history window size on the responsiveness and robustness of caching performance. We also show  that different base caching policies have {\it complementary} performance in different caching scenarios. (see Section~\ref{sec:window}.)

\item We then employ a Deep Recurrent Neural Network, specifically 
Long Short-Term Memory (LSTM), to mine the temporal locality in content popularity evolution at different time scales. LSTM-based time series analysis can be applied to the number of requests/viewers or the time interval till the next request. We show experimentally that LSTM, especially when combined with LFU, can often improve the performance of LFU and LRU, but is still not dominant. (detail in Section~\ref{sec:LSTM}.)

\item We present CEC, our DRL-based ensemble learning framework in Section~\ref{sec:CEC}. We propose to use  {\it virtual cache} to facilitate the parallel operations of ensemble caching policies. We solve the DRL {\it state-space-explosion} problem by using virtual hit ratios as its input features, and address the {\it delayed-reward} problem by emulating the theoretically optimal caching policy with future request oracle in the training phase.

\item Through extensive evaluations using two sets of real video request traces, we demonstrate that CEC not only can strategically select the most suitable caching policy in realtime to achieve higher hit ratio than any single policy, but also can improve the robustness against the unexpected popularity changes that a single policy is not designed for. CEC model can be well generalized to different cache scenarios with low computation overheads. It has great potential for  deployment in real edge cache networks. (detail in Section~\ref{sec:evaluation}.)
\end{enumerate}
Since video accounts for more than $70\%$ of fixed and mobile Internet traffic today~\cite{cisco2018cisco}, we focus on video-on-demand caching in the following analysis and experiments. But our algorithms can be applied to caching of other content.

\vspace{-0.05in}
\section{Related Work}
\label{sec:related}
There is a rich research literature on caching. Many traditional caching algorithms have been proposed in different contexts, such as RANDOM, LRU, First-In First-Out (FIFO), LFU and Greedy-Dual-Size-Frequency (GDSF)~\cite{Fofack2012,Melazzi2014,ChiaTaiChan2000, cherkasova1998improving}. In the classical caching studies, content requests are assumed to draw from a stationary popularity distribution, the so-called Independent Reference Model (IRM)~\cite{IRM,IRM1,IRM2}. Recent measurement studies have demonstrated that IRM cannot model the intrinsic non-stationarity in online traffic~\cite{li2016popularity,song2017learning}.
For better adaptation to the time-varying popularity patterns, forecast-based cache replacement policies have recently been proposed \cite{Ma2017,Zhang2019}. In~\cite{li2018data}, authors utilize Matrix Factorization (MF) based user interest forecast to enhance edge caching performance. There are also works on video content caching~\cite{poularakis2016} and mobile edge caching~\cite{Wang2017,Andre2017edgecache}. Deep Reinforcement Learning has also recently been applied to content caching area. For example,~\cite{somuyiwa2018reinforcement,zhong2018deep,Kirilin2020,fan2020pa,sadeghi2019reinforcement,sengupta2014learning,jiang2019multi,sadeghi2017optimal,Zhu2018,Wu2019,wang2020intelligent} use various critical features to train DRL models in an evolving manner so as to serve requests with popularity fluctuations and bursts. While most existing DRL caching policies generate caching decisions directly, our study focuses on DRL-based policy combination.




\vspace{-0.05in}
\section{Problem Description and Base Policies}
\label{sec:window}

\subsection{Caching under Dynamic Content Popularity}
The core idea of caching is to place the most popular items in the limited buffer space to maximally serve user requests. To keep track of content popularity evolution, the cached items are constantly updated in proactive and/or reactive manners. 
%
{\it The key to maximize caching gain is to accurately predict and keep track of the future content popularity}. From this perspective, the traditional algorithms use simple content request statistics to predict future popularity. For example, the rationale behind LRU is  that the more recently accessed content will be more popular in the near future, while LFU works under the assumption that the content accessed the most in the past will remain to be the most popular in future. When the user population handled by a cache server is large, such simple prediction can achieve good performance because of the stable content popularity distributions. 

In real content delivery systems, both content and users are highly dynamic: new items are constantly added to the catalog, user interests are fast changing, and the active population served by a cache box is time-varying, due to the temporal variations of user activities and user mobility. As a result, the content popularity presents strong and complicated temporal variations. This is particularly evident for edge caching, where the user population served by each cache box is small.    
%
%
A recent study~\cite{traverso2015unravelling,traverso2013temporal} on YouTube video requests  demonstrate that requests for video content have diverse long-term and short-term temporal locality. Shot noise model was proposed to model the content popularity, where the average request rate for content $c$ at time $t$ is modeled as $V_c  \lambda (t-\tau_c)$, where $\tau_c$ is the birth time for $c$ and $V_c$ is the total request volume, and $\lambda(\cdot)$ characterizes the temporal evolution of popularity. Different types of content have different life-spans and different popularity evolution patterns. For example, viral short videos on social media tend to have large volumes, but relatively short life-spans, while block-buster movies on Netflix can also have large volumes and long life-spans.

\begin{figure*}[htbp]
\centering{
 \begin{subfigure}[b]{.49\linewidth}
            \includegraphics[width=\linewidth]{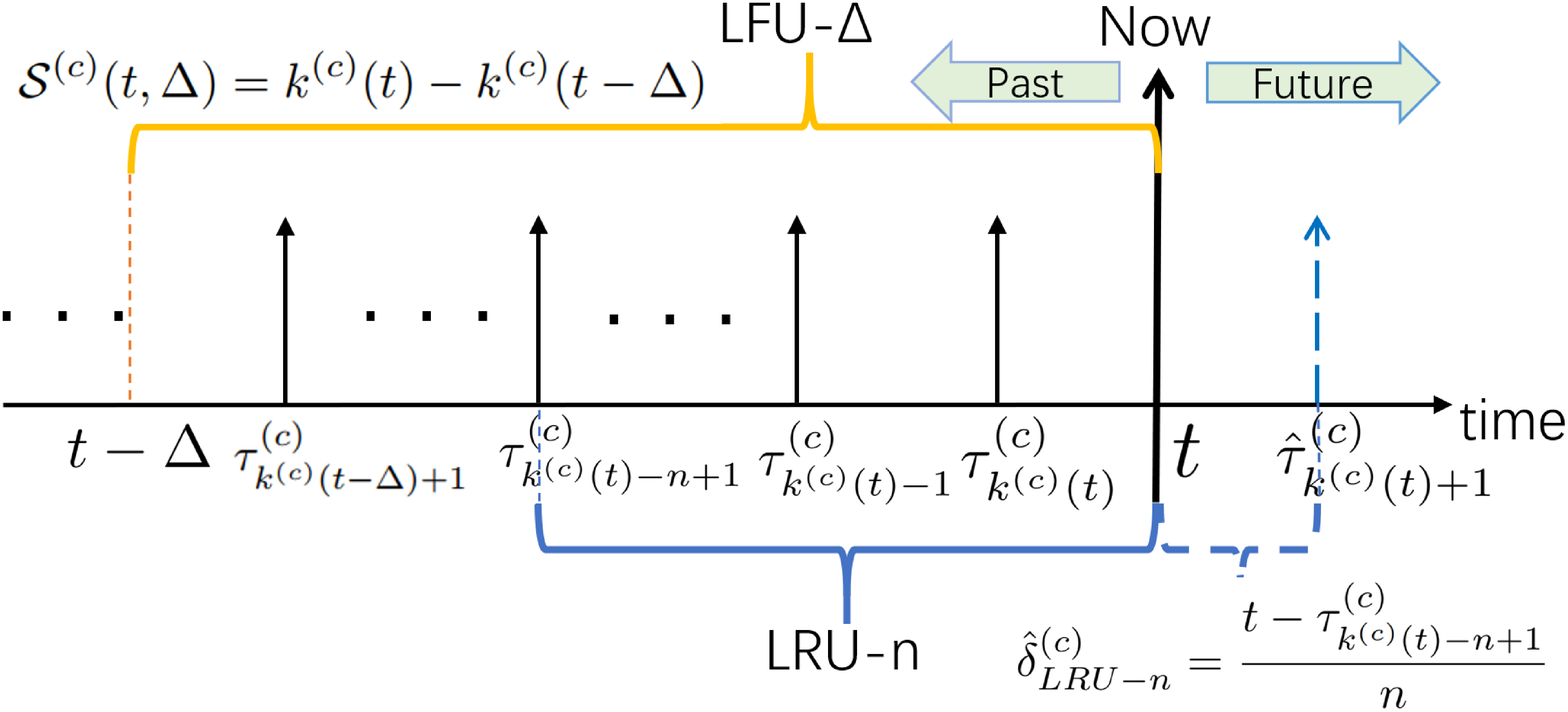}
            \label{fig:lfulru}
    \end{subfigure}
    \begin{subfigure}[b]{.49\linewidth}
            \includegraphics[width=\linewidth]{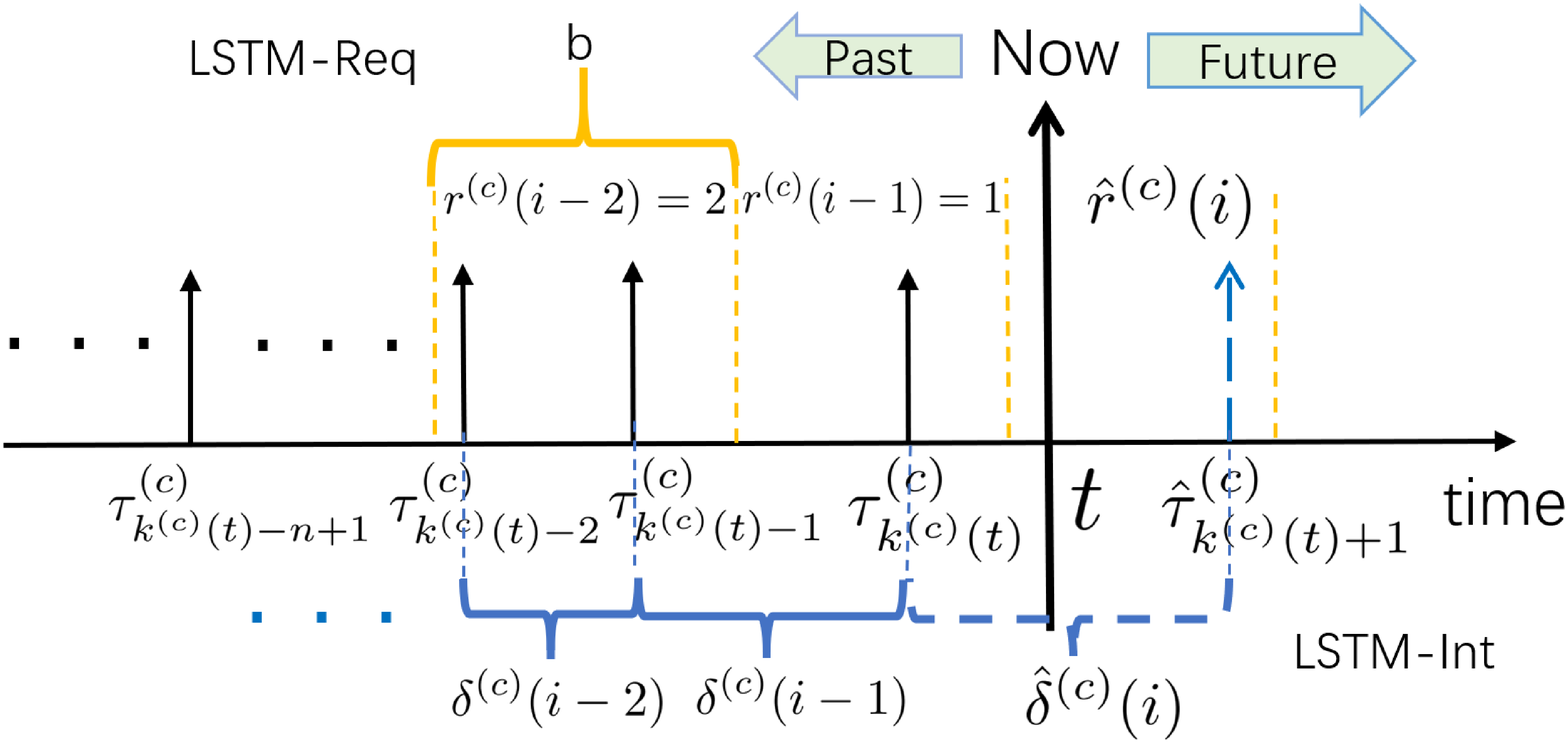}
            \label{fig:lstmintreq}
    \end{subfigure}

}
\caption{Time Series Analysis by LFU-$\Delta$, LRU-$n$, LSTM-Req and LSTM-Int}
\label{fig:lfulrulstm}
\end{figure*}

\subsection{Sliding-window based Caching Policies}
To serve content with highly dynamic popularity, caching algorithms have to learn and exploit the short-term and long-term temporal locality in  popularity evolution. Let $\mathcal C$ be the set of items, i.e., the video catalog. At any given time $t$, let $\mathcal R^{(c)}(t) \triangleq \{0 < \tau^{(c)}_1 \cdots \tau^{(c)}_i, \cdots \tau^{(c)}_{k^{(c)}(t)} \le t\}$ be the request history of $c \in \mathcal C$ up to $t$,  where $\tau^{(c)}_i$ is the arrival time of the $i$-th request out of $k^{(c)}(t)$ total requests for $c$ up to $t$. We need to estimate the future popularity of item $c$  based on the request history $\mathcal R^{(c)}(t)$. For the caching purpose, what really matters is the {\it relative ranking} of the estimated future popularities of all the items. Different caching algorithms calculate caching scores differently, and make proactive or reactive caching decisions based on caching scores.

\subsubsection{LFU-$\Delta$ Variants}
In the traditional LFU, the future popularity of $c$ is estimated as the total number of requests up to $t$. In other words, the caching score 
$\mathcal S^{(c)}(t)=k^{(c)}(t)$, and the item with the lowest caching score will be evicted. When the content popularity is stationary, let $\lambda^{(c)}$ be the average request rate for $c$, then $\mathcal S^{(c)}(t)$ converges to $\lambda^{(c)} t$ when $t$ is large. LFU score is proportional to the stationary popularity $\lambda^{(c)}$, and LFU is optimal under IRM. When the content popularity is highly dynamic, the expected request rate $\lambda^{(c)}(t)$ varies over time. The request patterns in distant history will not be predictive for content popularity in the near future. To deal with this, a natural fix is to use the number of requests within a sliding recent history window of $(t-\Delta, t]$, instead of the whole history, as the caching score: 
\begin{equation}
\label{eq:LFU-n}
\mathcal S^{(c)}(t,\Delta)=k^{(c)}(t) - k^{(c)}(t-\Delta).
\end{equation} 
The expected caching score becomes: 
\begin{equation}
\label{eq:LFU-n-E}
\E[\mathcal S^{(c)}(t,\Delta)]=\int_{t-\Delta}^t \lambda^{(c)}(x) dx,
\end{equation} 
which is proportional to the average request rate in $(t-\Delta, t]$. The illustration of LFU-$\Delta$ is shown in Fig.~\ref{fig:lfulrulstm}.  However, it is non-trivial to select the optimal $\Delta$.
\begin{itemize}
\item It is obvious that $\Delta$ should be dependent on how fast $\lambda^{(c)}(x)$ changes overtime. For fast-varying popularity, i.e. UGC with short lifespans, a small $\Delta$ will make caching respond quickly to popularity changes. {On the other hand, if $\Delta$ is too small, the caching algorithm will be mostly driven by the short-term popularity oscillations, and may lead to too frequent content replacement.}  In addition, with a small $\Delta$, the number of requests within $(t-\Delta, t]$ might be too small and the caching score calculated using those samples according to (\ref{eq:LFU-n}) is far way from its mean value in (\ref{eq:LFU-n-E}), therefore cannot serve as a reliable popularity estimator. This {\it data-sparsity} problem can be severe when the content request rate is low.   
\item What is less obvious is that the choice of  $\Delta$ should also consider the cache size. When the cache size is larger, a content tends to stay in the cache for a longer time period. As a result, the ``gain" of caching $c$ at time $t$ is the expected total number of hits of $c$ in a longer future time horizon till its eviction. {\it It makes good sense to estimate the caching gain for a larger cache size using larger history window size $\Delta$.} \end{itemize}



\begin{table*}[htb]%
\centering
\caption{Hit Ratios of LFU-$\Delta$ on Nine Different Caching Scenarios}
\label{table:LFU example}
{
\begin{tabular}{||c||c|c|c||c|c|c||c|c|c||}
\hline

\diagbox{}{Scenarios}  & \multicolumn{3}{c||}{\bfseries Cache Size= 50} & \multicolumn{3}{c||}{\bfseries Cache Size=100} & \multicolumn{3}{c||}{\bfseries Cache Size=500} \\
  
  \cline{2-10}

 $\Delta$ (requests)   & \bfseries Trace 1 & \bfseries Trace 2 & \bfseries Trace 3    & \bfseries Trace 1  &  \bfseries Trace 2     &  \bfseries Trace 3   & \bfseries Trace 1         & \bfseries Trace 2     &\bfseries Trace 3   \\
\hline
500     &\bf{0.0858}   & \bf{0.0905}       &\bf{0.0737}     & \bf{0.1115}       &  0.1125           & 0.0917        & 0.1906        & 0.1919            &  0.1674   \\
5000    & 0.0660       &  0.0792           & 0.0722         & 0.0981            & {\bf 0.1132}      & 0.1044        & 0.2281        & 0.2395            &  0.2216   \\
50000   & 0.0616       &  0.0745           & 0.0714         & 0.0938            &  0.1070           &\bf{0.1045}    & \bf{0.2399}   & \bf{0.2531}       & {\bf 0.2483}   \\ 
$\infty$     & 0.055        &  0.0655           & 0.0623         & 0.0864            &  0.0985           &0.095          & 0.2333        & 0.2441            &  0.2378   \\\hline
\end{tabular}
}
\end{table*}

\begin{table*}[htb]%
\centering
\caption{Hit Ratios of LRU-n on Nine Different Caching Scenarios}
  \label{table:LRU example}
{
\begin{tabular}{||c||c|c|c||c|c|c||c|c|c||}
\hline

\diagbox{}{Scenarios}  & \multicolumn{3}{c||}{\bfseries Cache Size=50} & \multicolumn{3}{c||}{\bfseries Cache Size=100} & \multicolumn{3}{c||}{\bfseries Cache Size=500} \\
  
  \cline{2-10}

 $n$  & \bfseries Trace 1 & \bfseries Trace 2 & \bfseries Trace 3    & \bfseries Trace 1  &  \bfseries Trace 2     &  \bfseries Trace 3   & \bfseries Trace 1         & \bfseries Trace 2     &\bfseries Trace 3   \\
\hline
1  & \bf{0.0808}    & 0.0754       & 0.0577         & \bf{0.1018}   & 0.0985        & 0.0786        & 0.1925        & 0.1944        &  0.1694\\
2  & 0.0701         & \bf{0.0803}  &  0.0662        & 0.0953        &  0.1076       & 0.0934        & 0.2143        & 0.2291        & 0.2127   \\
4  & 0.0625         & 0.0777       & 0.0675         & 0.0924        & {\bf 0.1086}  & 0.0991        & 0.2283        & 0.2444        & 0.2352   \\
8  & 0.0629         & 0.0755       & {\bf 0.0708}   & 0.0947        & 0.1079        & \bf{0.1036}   & \bf{0.2380}   & \bf{0.2512}   & {\bf 0.2453}\\ \hline
\end{tabular}

}

\end{table*}

Using real content request traces (more details in Section~\ref{sec:evaluation}), we study LFU-$\Delta$'s performance on different caching scenarios in Table \ref{table:LFU example}. History window size is measured by the number of past requests within the window. {Trace 1, 2, 3 are requests generated over four days by users from three subnets with different sizes. Each subnet can be served by a cache with different buffer sizes of 50, 100, 500 videos~\footnote{To focus on content popularity, we made the simplifying assumption that all videos have the same size. In reality, videos with different sizes can be stored in the units of chunks with the same size. And each video request can be converted into multiple chunk requests, based on the video size.}. Each column is the hit ratios of different window sizes for a trace-cache size combination. The best performance is marked with bold fonts. }
{\it It is clear that window-based LFU is better than the traditional LFU ($\Delta=\infty$), and no single window size dominates in all scenarios.} It is difficult to come up with an optimal rule for tuning $\Delta$ without an accurate dynamic model of popularity evolution. This motivates our model-free DRL approach that adapts the history window of base caching policies, including LFU, to optimize the final hit ratio.

\subsubsection{LRU-n Variants}
For reactive caching with the ``oracle" of complete future request arrivals, the sample-path-wise optimal reactive caching strategy has been shown to be the Farthest-in-Future (FIF) algorithm that evicts the content in the cache that is not requested until the farthest in the future~\cite{Belady1966}. In other words, the FIF caching score is calculated as  
\begin{equation}
\label{eq:fif_score}
S^{(c)}(t)=t-\tau^{(c)}_{k^{(c)}(t)+1}.
\end{equation}
FIF evicts the one with the lowest score (largest $\tau^{(c)}_{k^{(c)}(t)+1}$). In reality, the arrival time of the next request $\tau^{(c)}_{k^{(c)}(t)+1}$ is not available, but can be estimated using different algorithms.  In the traditional LRU, the time till the next request arrival is simply estimated as the time elapsed since the last arrival:  
\[\hat \tau^{(c)}_{k^{(c)}(t)+1} -t = t-\tau^{(c)}_{k^{(c)}(t)}.\]
Accordingly, the caching score is calculated as $\mathcal S^{(c)}(t)=-(t-\tau^{(c)}_{k^{(c)}(t)})$,  The least-recently requested content will be evicted. If the content arrival process is Poisson with constant rate $\lambda^{(c)}$, due to the memoryless property of Poisson process, at any given time $t$, the time elapsed since the last arrival follows the same exponential distribution with rate $\lambda^{(c)}$ as the time till the next arrival. So  $t-\tau^{(c)}_{k^{(c)}(t)}$ is an unbiased estimator for $\tau^{(c)}_{k^{(c)}(t)+1}-t$. And the expected caching score becomes $\E (\mathcal S^{(c)}(t)) = - \frac 1 {\lambda^{(c)}}$. Similar to LFU, LRU score increases with stationary content popularity $\lambda^{(c)}$, and LRU is also optimal under IRM in the expected sense.

However, the vanilla LRU score only takes into account the timing of the previous one request of each item, which cannot serve as a reliable estimator. To address this issue, one can calculate caching scores based on the inter-arrivals of the previous $n$ requests:
\begin{align}
\label{eq:LRU-n}
\mathcal S^{(c)}(t,n) &= t-\tau^{(c)}_{k^{(c)}(t)}+\sum_{i=0}^{n-2} \left(\tau^{(c)}_{k^{(c)}(t)-i}-\tau^{(c)}_{k^{(c)}(t)-i-1}\right) \nonumber\\
&=t-\tau_{k^{(c)}(t)-n+1}.
\end{align}
For stationary Poisson arrival with constant rate $\lambda^{(c)}$, the expected LRU-n score becomes $\E [\mathcal S^{(c)}(t,n)] = - \frac n {\lambda^{(c)}}$. The score based on $n$ i.i.d inter-arrival samples is a more reliable popularity indicator than the traditional LRU score based on a single sample. For requests with time-varying popularity $\lambda^{(c)}(t)$, the expected LRU-n score is: 
\begin{equation}
\label{eq:LRU-n-E}
\E [\mathcal S^{(c)}(t,n)] = - \sum_{i=0}^{n-1} \frac 1 {\lambda^{(c)}\left(\tau^{(c)}_{k^{(c)}(t)-i}\right) },
\end{equation}
which is inversely proportional to the negation of the harmonic mean of the expected arrival rate at the previous $n$ arrivals. Similar to the window size tradeoff in LFU-$\Delta$, the number of previous requests to be considered by LRU-$n$ tradeoffs the responsiveness to popularity changes and continuity/stability of content placement. The illustration of LRU-$n$ is shown in Fig.~\ref{fig:lfulrulstm}. 
LRU-$n$'s performance in the nine scenarios of Table~\ref{table:LFU example} are shown in Table \ref{table:LRU example}. Again, no single $n$ value dominates.

\begin{table*}[htb]%
\centering
\caption{Comparison between LFU, LRU and LSTM Policies on Different Caching Scenarios.}
  \label{table:comparison example2}
{
\begin{tabular}{||c||c|c|c||c|c|c||c|c|c||}
\hline

\diagbox{}{Scenarios}  & \multicolumn{3}{c||}{\bfseries Cache Size=50} & \multicolumn{3}{c||}{\bfseries Cache Size=100} & \multicolumn{3}{c||}{\bfseries Cache Size=500} \\
  
  \cline{2-10}

 Method  & \bfseries Trace 1\textprime & \bfseries Trace 2\textprime & \bfseries Trace 3\textprime    & \bfseries Trace 1\textprime  &  \bfseries Trace 2\textprime     &  \bfseries Trace 3\textprime   & \bfseries Trace 1\textprime         & \bfseries Trace 2\textprime     &\bfseries Trace 3\textprime   \\
\hline
LFU-$\Delta^\ast$  & \bf{0.0779}    & \bf{0.0913}       & 0.0725         & 0.1002   & 0.1173        & 0.1065        & 0.2425        & 0.2660        &  \bf{0.2490}\\
LRU-$n^\ast$  & 0.0700         & 0.0813  &  0.0708        & 0.0912        &  0.1110       & 0.1040        & 0.2446        & 0.2668        & 0.2474   \\
LSTM-Req($b=10$)  & 0.0508         & 0.0721       &  0.0692         & 0.0726        & 0.0956  & 0.1007        & 0.2034        & 0.2265        & 0.2268   \\
LSTM-Req($b=5$) &0.0480    & 0.0460  &0.0660   &0.0716 & 0.0649  & 0.0940  &0.1855 &0.1769   &0.1992\\
LSTM-Int  &  0.0748                & 0.0912          & {\bf 0.0755}   & \bf{0.1108}        &  \bf{0.1272}       & \bf{0.1122}   &\bf{0.2480}    &\bf{0.2702}    & 0.2480\\ \hline
\end{tabular}

}
\end{table*}

\section{LSTM-based Caching Policies}
\label{sec:LSTM}
The sliding-window based caching algorithms presented in the previous section aim at estimating the future content popularity using requests within a finite time/request window. However,  all requests within the window have the same importance weights in summation based estimations (\ref{eq:LFU-n-E}) and (\ref{eq:LRU-n-E}), and request history outside of the window is totally ignored.  


\subsection{Time Series Analysis}
To fully explore the temporal locality of popularity evolution at different time scales, we need a more flexible prediction framework that can adaptively choose the history window size and determine the relative importances of history data to maximize the prediction accuracy.
We now study future content popularity prediction as a time-series analysis problem. Specifically, given the content request history, $\mathcal R^{(c)}(t)$, we can derive different types of time series $\{x^{(c)}(i)\}$ and predict the future values of $x^{(c)}(i)$ to calculate the caching score for $c$. 
\subsection{LSTM-based Popularity Prediction}
There are many methods for time series prediction, ranging from simple averaging adopted by LFU-$\Delta$ and LRU-$n$,  signal processing approaches, such as Kalman filter~\cite{brown1992introduction} and Recursive least squares (RLS)~\cite{haykin2008adaptive}, to more sophisticated machine learning methods, in particular Recurrent Neural Network (RNN) for sequence model. For a dynamic system with input $\mathbf{x(t)}$, RNN models the system evolution with hidden state $\mathbf {h(t)}$ using a neural network parameterized by $\theta$:
\[\mathbf{h(t)}=f(\mathbf{x(t)}, \mathbf{h(t-1)}; \theta).\]
Due to the recurrent structure, system state $\mathbf{h(t)}$ contains information of the whole sequence up to $t$. The output will be generated based on the system state: $\mathbf{y(t)}=g(\mathbf{h(t)}; \phi)$. In principle, RNNs can exploit arbitrary long-term dependencies in the input sequences to improve prediction accuracy.  However, the traditional RNNs are vulnerable to the ``vanishing gradient" problem, where the gradients vanish when back-propagating for many steps.  LSTM and GRU~\cite{gers1999learning} were introduced to partially solve the problem by introducing gate functions to control the information flow within RNN. Through training, LSTM/GRU can learn the correlation structures with the strongest prediction power for the target output. LSTM/GRU have become the state-of-the-art tools in sequence modeling, e.g. natural language model and time series analysis.  

\subsubsection{{\bf LSTM-Int} -- Estimating the Arrival Time of the Next Request}
 Following the direction of emulating the optimal FIF caching strategy, we use LSTM to estimate the arrival time of the next request. Similar to LRU-$n$, we use LSTM to process the time series of content request inter-arrivals $\{ \delta^{(c)}(i)\}$. As is shown in Fig.~\ref{fig:lfulrulstm}, the input of LSTM is the vector of the previous $w$ inter-arrivals, $\mathbf {x^{(c)}(i)}=\{\delta^{(c)}(i-1), \cdots, \delta^{(c)}(i-w)\}$, the output is the scalar of the next inter-arrival  $\mathbf{y^{(c)}(i)}=\delta^{(c)}(i)$. With the predicted next inter-arrival time  $\hat \delta^{(c)}_{k^{(c)}(t)+1}$, the caching score is calculated as 
\begin{equation}
\label{eq:LSTM2}
\mathcal S^{(c)}(t) = t - \hat \delta^{(c)}_{k^{(c)}(t)+1} -  \tau^{(c)}_{k^{(c)}(t)}.
\end{equation}
While $w$ determines the number of past samples going to the input vector, since LSTM memorizes history information using its internal memory cell, it can capture temporal correlation between $\delta^{(c)}(i)$ even if $w=1$. We set $w=2$ in experiments.  

    

    

\subsubsection{{\bf LSTM-Req} -- Estimating the Number of Requests in the Next Slot}
Similar to LFU-$\Delta$, we use LSTM to process the time series of the number of requests within each time slot $\{ r^{(c)}(i)\}$. As is shown in Fig.~\ref{fig:lfulrulstm}, the input of LSTM is the vector of request numbers in the previous $w$ slots, 
$\mathbf {x^{(c)}(i)}=\{r^{(c)}(i-1), \cdots, r^{(c)}(i-w)\}$, 
the output is the scalar of the number of requests in the next slot,  $\mathbf{y^{(c)}(i)}=r^{(c)}(i)$. The predicted request number in future time slot $(t, t+b]$ is directly used as the caching score.   Similar to LSTM-Int, $w$ is not a critical parameter to tune. On the other hand, the time slot length $b$ controls the time-resolution of the time series fed into LSTM, as well as the prediction time horizon. It is an important parameter to tune, which is shown in Table \ref{table:comparison example2}. When the view duration of each request is also available, one can generate the time series of the number of active viewers, then LSTM can predict the number of viewers in the next time slot similar to {\bf LSTM-Req}.



\subsection{Caching Implementation and Performance}
The proposed LSTM models will be trained offline using respective time series derived from request history. The trained LSTM model will be run online to generate caching scores for items in realtime. For LSTM-Int, upon a new request for item $c$, the trained LSTM model will be called to predict the arrival time of the next request for $c$ and update its caching score, which will be used for caching replacement. The LSTM online inference time is at millisecond level.  For LSTM-Req, at the beginning of each time slot, the trained LSTM model will be called once for each active item. For realtime operation, the time slot can not be too short. In our experiments, when $b>1$ minute, the LSTM inference can be done in realtime. 

Due to the heterogeneity in popularity, the range of the request inter-arrival time $\delta$ is quite large. If we directly use the $\delta$ values as the input and output labels of LSTM, the training loss will be dominated by the large $\delta$ of items that are most likely not cacheable. To make LSTM model focus on accurately predicting inter-arrival values of cacheable items, we convert each $\delta$ value into a categorical label. Specifically, based on the CDF of $\delta$ in the training set, we divide the value range into $P$ partitions, and the boundaries of the partitions are set so that the fraction of $\delta$ samples following into each partition is $1/P$. We set $P=16$ in experiments. {Then we convert each $\delta$ value into an one-hot vector corresponding to the partition that it falls into. Similarly, each output label is the corresponding one-hot vector of the actual inter-arrival value.} The goal of LSTM is to accurately predict the right partition the next inter-arrival belongs to. The training loss function is categorical cross-entropy.

For cold-start items without enough number of past requests, LSTM-Int cannot generate caching scores for them. In addition, if the predicted arrival time for the next request has passed for a while, but the request has not arrived yet, the caching score calculated by LSTM-Int prediction becomes invalid. Finally, for large predicted $\delta$ values, their relative ranking becomes less reliable than small values \footnote{The partitions at the upper range are much wider than the partitions at the lower range.}. For all the aforementioned cases, we revert back to LFU-$\Delta$  to calculate caching score. And content without valid LSTM-Int scores will be evicted based on their LFU-$\Delta$ first. {\it LSTM-Int is indeed a combined policy between LSTM and LFU-$\Delta$.} 

In Table~\ref{table:comparison example2},  we compare the performance of LFU, LRU and LSTM based policies on the nine scenarios used in Table~\ref{table:LFU example} and \ref{table:LRU example}. For each trace, we report the average hit ratio of the first out of the four days. For each scenario, we pick the best configuration for LFU and LRU according to Table~\ref{table:LFU example} and \ref{table:LRU example}. For LSTM-Req, we tried time slot lengths of $5$ and $10$ minutes. LSTM-Int dominates LSTM-Req with different $b$ values, and it is the best-performing policy for six out of nine scenarios. LFU with optimal window configuration is the best for the other three scenarios. Due to the better performance and lower computation overhead of LSTM, we use LSTM-Int policy to represent LSTM based policies in the following.  
\section{DRL-based Cocktail Edge Caching}
\label{sec:CEC}
We now present our Deep Reinforcement Learning (DRL) based {\it cocktail edge caching} framework that adaptively selects one out of an ensemble of caching policies that is most suitable for handling the current content mix and  popularity trends. 

\subsection{DRL for Caching}
A cache processes content requests sequentially. There is strong temporal correlation between consecutive caching decisions. Given the content and user dynamics, optimal caching can be studied as a {\it stochastic optimal dynamic control} problem. One direction is to develop dynamic system models, with the cached content as its state, and state transition probability determined by the content popularity profile following the IRM model~\cite{IRM}, and construct optimal caching policy to maximize the long-term cache hit ratio. However, if the content popularity is not stationary, the obtained caching policy will not achieve the optimal performance~\cite{li2018accurate}. 

It is therefore more promising to investigate dynamic caching under the model-free optimal control framework known as Reinforcement Learning (RL). Through ``trial-and-error" over a wide range of content request traces with different content mix and popularity evolution patterns, a RL agent can be trained to generate good caching policies that generalize well in real operations. Deep Reinforcement Learning (DRL) uses deep neural networks (DNN) to generate action policy, which is more adaptive and extensible to complicated problems than the traditional RL. DRL has recently been applied to content caching, especially edge caching~\cite{somuyiwa2018reinforcement,sadeghi2019reinforcement,zhong2018deep}. The common challenge is to deal with the huge catalog size in real systems.  A DRL algorithm is characterized by its $\langle \texttt{STATE}, \texttt{ACTION}, \texttt{REWARD} \rangle$ tuple.
A naively-designed DRL caching agent would take the complete request history of all items $\{\mathcal R^{(c)}(t), c \in \mathcal C \}$ as its \texttt{STATE},  and generate policies for \texttt{ACTION} on cache replacement to maximize the \texttt{REWARD} of cache hit ratio. However the state space dimension is $\Theta(|\mathcal R^{(c)}(t)|^{|\mathcal C|})$. Any reasonable catalog size $|\mathcal C|$ will lead to the {\it state space explosion} problem.  

We employ a two-level hierarchy to address this challenge. The lower level is an ensemble of caching policies which process the content requests and make caching decisions in parallel. The upper level DRL agent does not directly process requests, nor generate caching  decisions. Instead, it monitors the performance of the low-level policies as well as summary statistics of current request patterns. Its action is to dynamically select the caching policy that is the most suitable to the current situation to control the 
cache. The DRL agent is essentially an {\it adaptive policy selector}. This design makes the state space of DRL agent 
well-manageable, and can work with any cache system with arbitrary configuration and request pattern. The DRL agent is trained to combine the merits of constituent caching algorithms to address the complicated content mix and dynamic popularity trends. The high-level idea is similar to the classic combination therapy for AIDS, namely AIDS cocktail~\cite{henkel1999attacking}. We call our framework {\bf Cocktail Edge Caching (CEC).} 

\subsection{Policy Ensemble and Virtual Cache}

Definition \textit{1:}\hspace{3mm}\textbf{Policy Ensemble} $\mathcal E$ is an ensemble of caching policies pre-selected based on their performance on historical request traces. Some may work the best on certain traces, while the others may work the best on other traces. No policy is dominated by any other single policy on all traces. Let $\{p_1, p_2, ..., p_{|\mathcal E|}\}$ denote the pre-selected policies in $\mathcal E$.

Definition \textit{2:}\hspace{3mm}\textbf{Primary Cache} $PC$ is the real cache used for storing the items. The caching performance is measured by the {\it hit ratio} of all the content requests.  

To select the best policy, the DRL agent needs to know the performance of all policies on the recent requests. Since only one policy is selected to control the primary cache at any given time, we set up {\bf Virtual Caches}, one for each constituent policy, to evaluate their performance. A virtual cache does not have any buffer space to store real content. It only maintains a simple dictionary that stores the identities of the virtually cached items. A similar concept, called Ghost cache, was proposed in~\cite{megiddo2003arc}, where cache lists of ghost caches are combined to form the primary cache list. We only use virtual caches to evaluate the performance of policies, and the primary cache is controlled by one policy at any given time.

Whenever the primary cache receives a request, it broadcasts the requested content ID to all the virtual caches, each of which will check whether the ID is in its virtual cache list: if yes, a {\it virtual hit} is logged; if not, a {\it virtual miss} is logged, and some ID currently in the list will be evicted based on its associated caching policy. Each cache periodically calculates and reports its {\it virtual hit ratio} to the DRL agent.  All virtual caches receive the same request sequence as the primary cache. Due to policy difference, the virtual cache lists will be different from each other and the primary cache list. To control the divergence, all virtual cache lists will be synchronized to the primary cache list after certain number of requests.  This is another main difference from the traditional ghost caches~\cite{megiddo2003arc}.

\subsection{DRL-based Policy Selection}

As mentioned above, different policies dominate under different scenarios, and it's hard to tell exactly what factors lead to the dominance of certain policy under certain scenario. Due to the temporal locality in content popularity, the recent performance of a policy can be used as an indicator for its performance in the near future. Let $\{h_1^q(t), h_2^q(t), ..., h_{|\mathcal E|}^q(t)\}$ denote the virtual hit ratios of all the policies within time window $(t-q,t]$.
 However, similar to the history window size selection problem for LFU-$\Delta$, we don't know exactly how far we should look back, and how to assign different weights to hit ratios within different mini-slots of the history window. We  resort to a DRL agent to learn how to select policy.

Other than the virtual performance of the virtual caches, we also feed some summary context information to the DRL agent to help its selection. First of all, recent \textbf{content popularity variations} can definitely affect all policies' performance and their relative ranking.  For example, when some newly generated items suddenly become popular, LFU reacts slower to the sudden popularity surge, and performs worse than LFU-$n$ and LRU. But items with rather stable popularity are definitely favorable to LFU. On the other hand, as shown in Table~\ref{table:LFU example} and \ref{table:LRU example}, \textbf{the total request volume} also has a significant influence on the relative ranking of policies. Therefore, the two factors are also used as state features for the DRL agent.
 
\subsection{DRL Agent Development}\label{AA}

\begin{figure}[htbp]
\centerline{\includegraphics[width=0.8\linewidth]{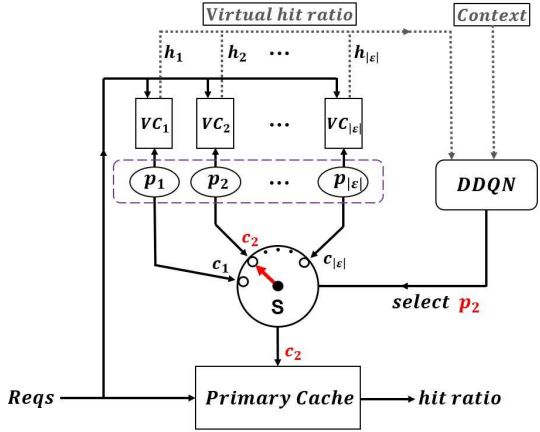}}
\caption{CEC Architecture}
\label{fig:drl}
\end{figure}

Fig.~\ref{fig:drl} is the architecture of CEC system.

\noindent\textbf{\texttt{Architecture:}} We use Double DQN (DDQN)~\cite{van2016deep}, a classic DRL algorithm, to train the policy selector. The traditional Deep Queue Network (DQN) learns the  long-term reward function $Q(S,a)$ for taking action $a$ under state $S$. The optimal action policy can be obtained as $\pi(S)=argmax_a Q(S,a)$. The Q function is approximated by a deep neural network (Q-network) parameterized by  $\theta$, which is updated in training to minimize the difference between the target Q-value and the estimated Q-value. DDQN is an extension of the  traditional DQN. It not only reduces the overestimation of DQN, but also performs better than DQN on several test platforms~\cite{van2016deep}. DDQN employs two DNNs, i.e., online Q-network and target Q-network. The online Q-network approximates the Q-value, and is updated by each batch of training samples. The target Q-network has the same structure as the online 
Q-network, but is updated less frequently, which makes better estimation of the target Q-value and stabilizes the training process.

\noindent\textbf{\texttt{STATE $S$:}} 
As mentioned above, the state features include the virtual hit ratios of each policy in the previous $10$ time slots, with each time slot consisting of $100$ requests. To measure the popularity variations, we calculate the number of overlapped items between the top-100 most frequently-requested items of the previous two request windows ($1,000$ requests within each window).  To summarize user activity level, we calculate the total number of requests for all items in the previous five minutes. The size of the state space is $10\times|\mathcal E|+2$. The temporal patterns in content popularity evolution will translate into the temporal patterns of caching performance. To fully explore such patterns, similar to LSTM-Int, the virtual hit ratio vector is first processed by a LSTM layer before being fed into the following fully-connected layers with the two context features. The outputs of the following layer are Q-values.

\noindent\textbf{\texttt{ACTION $a$:}} DRL agent's action is selecting one policy from $\mathcal E$. The action space size is  $|\mathcal E|$.  The selected policy controls the primary cache to decide which item to evict.  If the active caching policy is changed for each new request, the primary cache will become unstable. In both training and testing, policy selector is called once every $n$ requests. The selected policy controls primary cache for the next $n$ requests.   

\noindent\textbf{\texttt{REWARD $r$:}} In the training stage, the DRL agent is reinforced by the reward that its current action received from the environment.  Ideally, the  \texttt{REWARD} function should give accurate and timely feedback about the quality of the action. The natural choice is to use the actual hit ratio of the selected policy on the primary cache as its reward. However, as stated earlier, the performance impact of adding/evicting an item from cache can be fully observable after many future requests, the {\it delayed-reward} problem, which will significantly slow down the convergence and the solution quality of DRL. Fortunately, in the training phase, we have the ``oracle" of all future request arrivals of all items, and can use the optimal FIF caching strategy as the benchmark to evaluate the quality of policy selection. Specifically, for a policy $p_i$, if the item it evicts is $c_i$, we calculate $c_i$'s FIF score, i.e., the negation of the time till its next request, according to (\ref{eq:fif_score}). Since FIF always evicts the item with the lowest FIF score, to emulate FIF, the DRL agent should select the policy whose evicted item has the lowest FIF score among all the policies. For each request, we rank all policies in the increasing order of the FIF scores of their eviction candidates and sequentially assign policy scores $\{|\mathcal E|-1, |\mathcal E|-2, ..., 0\}$ to the sorted list. If there is a tie between two policies, they get the same  policy score. The reward for selecting $p_i$ as the active policy is calculated as the sum of $p_i$'s policy scores for the next $n$ requests.

\begin{figure*}[htbp]
\centering{
 \begin{subfigure}[b]{.32\linewidth}
            \includegraphics[width=\linewidth]{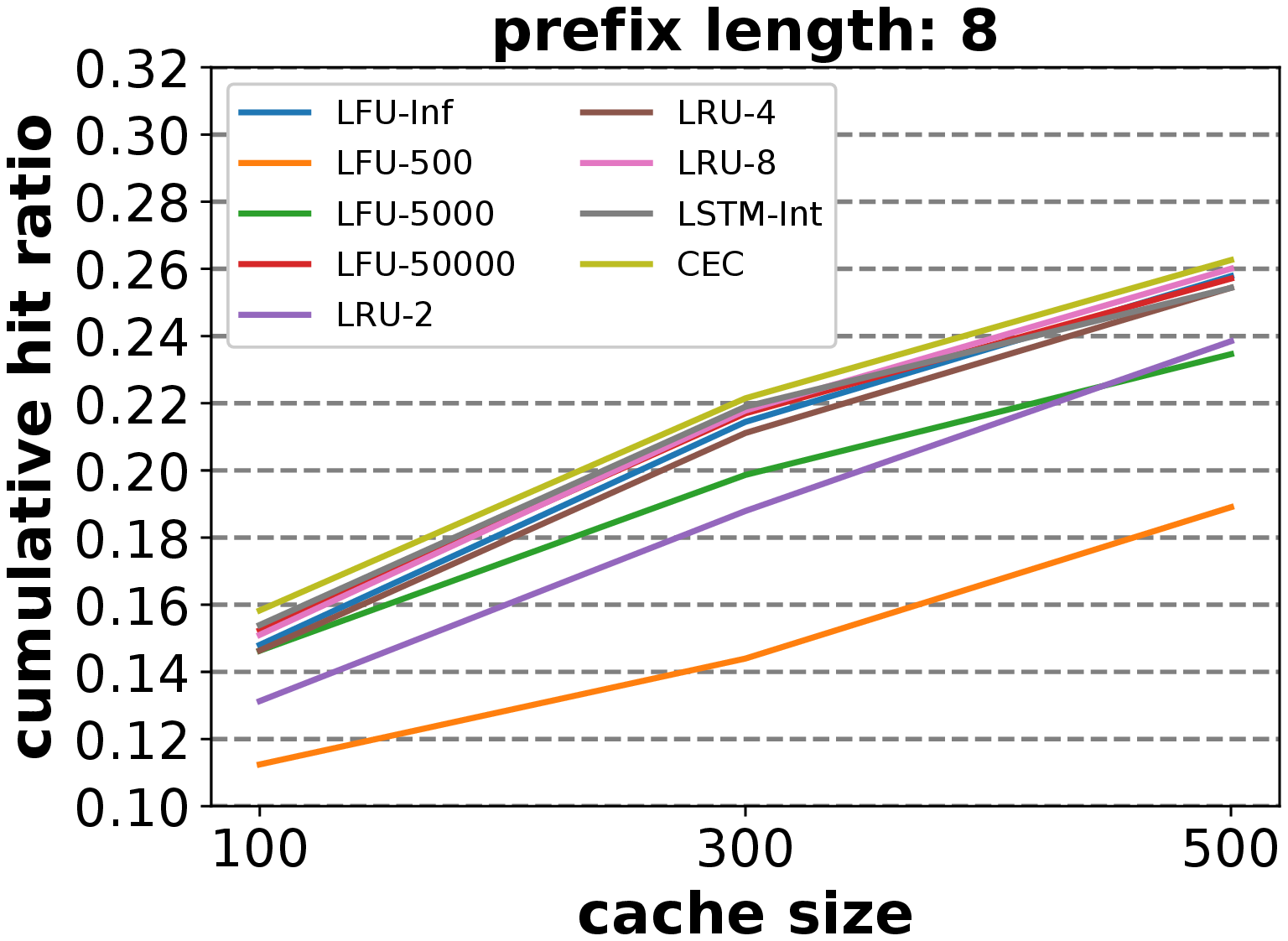}
    \end{subfigure}
    \begin{subfigure}[b]{.32\linewidth}
            \includegraphics[width=\linewidth]{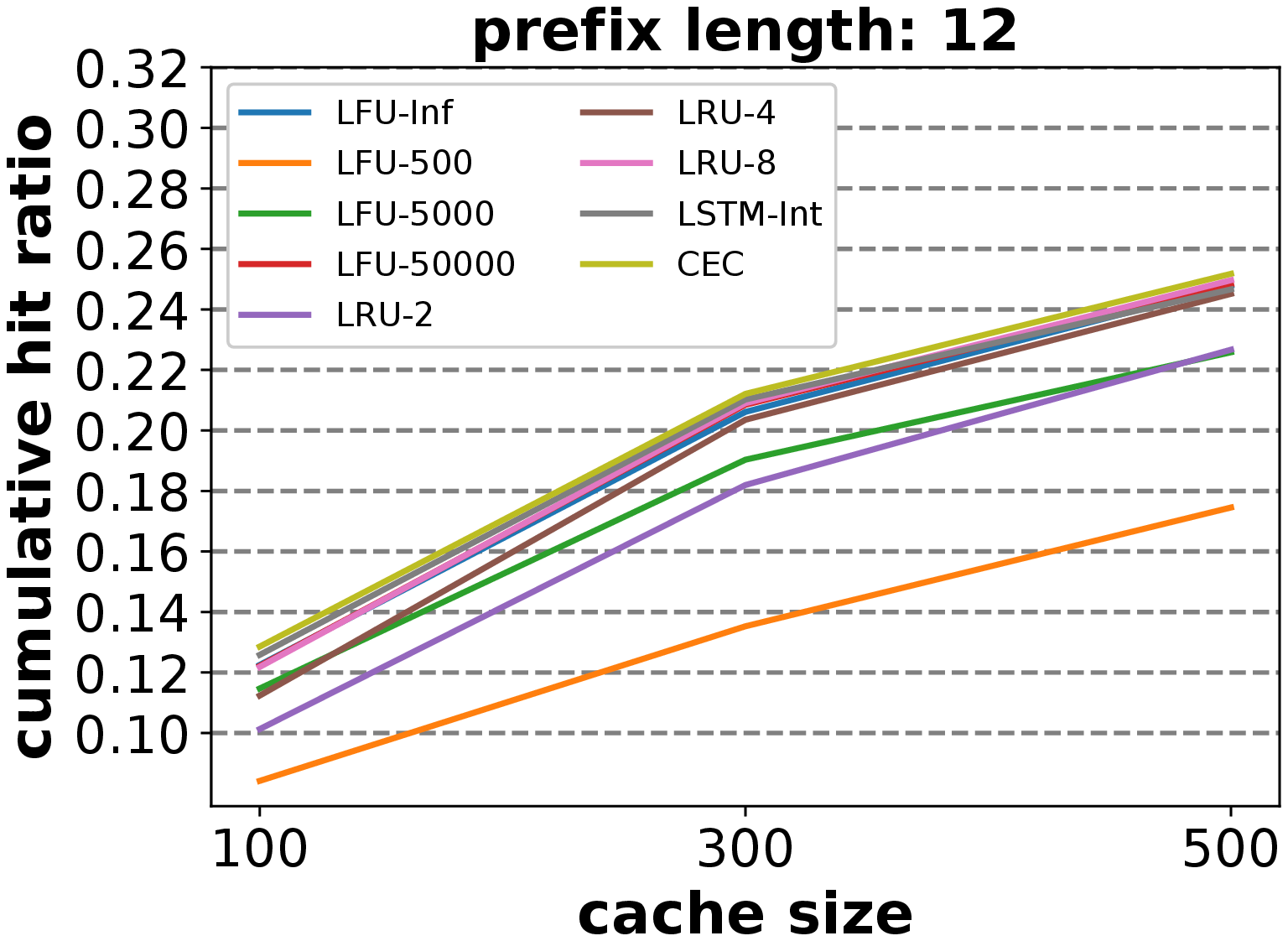}
    \end{subfigure}
    \begin{subfigure}[b]{.32\linewidth}
            \includegraphics[width=\linewidth]{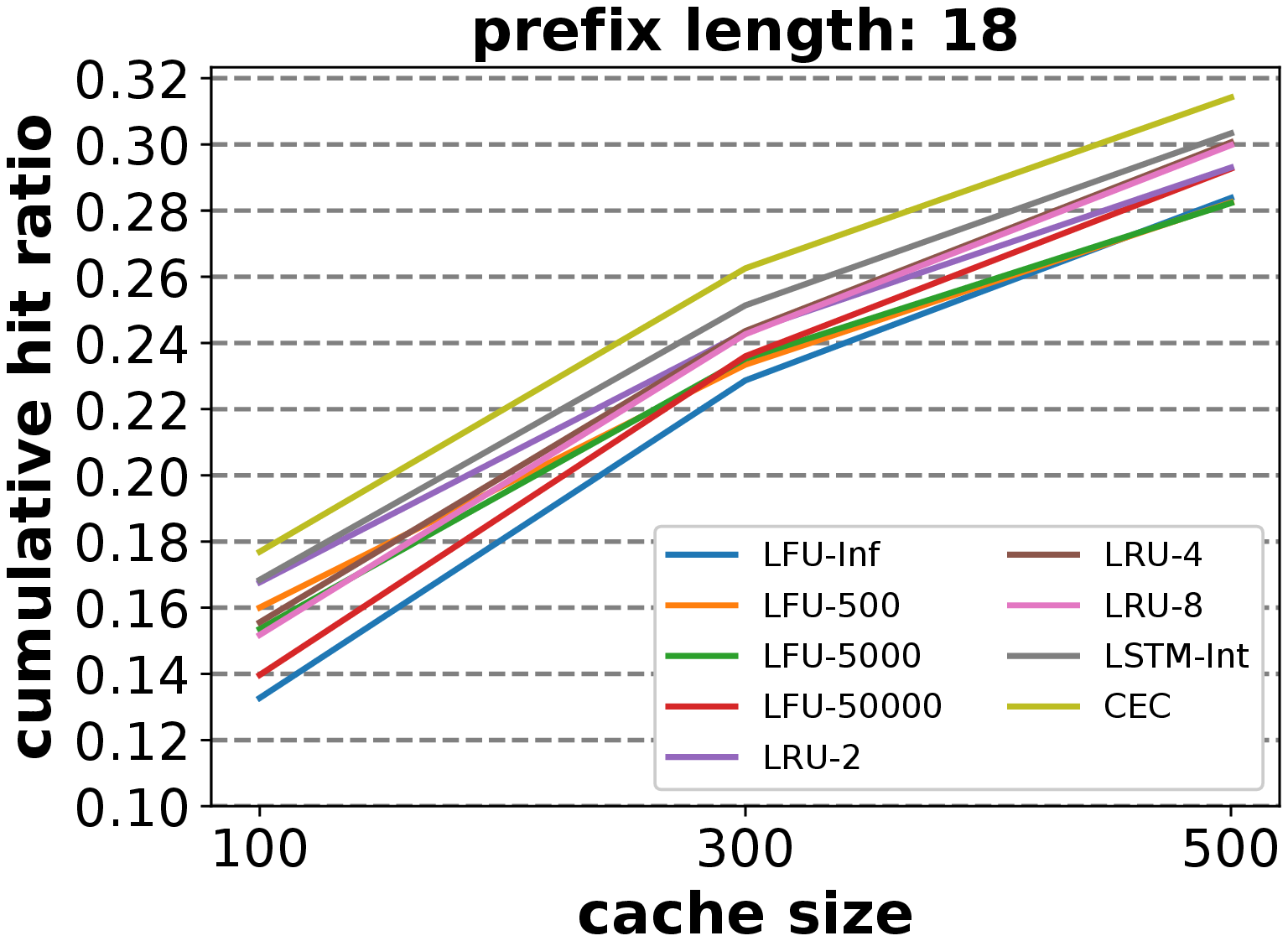}
    \end{subfigure}
}
\caption{\textbf{Cumulative Hit Ratio vs. Cache Size} on dataset A at different prefix lengths}
\label{fig:hr_cachesize}
\end{figure*}

\begin{table*}[htb]%
\centering
\caption{Relative Hit Ratio Improvement of CEC over Base Policies on Scales of Subnets for Dataset A}
  \label{table:relative_increase}
{
\begin{tabular}{||c||c|c|c|c|c|c|c|c||}
\hline

\diagbox{}{Baselines}  & & & & & & & & \\
  

 Prefix length  & \bfseries LFU-Inf & \bfseries LFU-500 & \bfseries LFU-5000    & \bfseries LFU-50000  &  \bfseries LRU-2     &  \bfseries LRU-4   & \bfseries LRU-8         & \bfseries LSTM-Int       \\
\hline
8  & 3.40\%    & 45.79\%       & 11.10\%         & 2.40\%   & 15.28\%        & 4.81\%        & 1.96\%        & 2.35\%        \\
12  & 2.73\%         & \textbf{51.59\%}  &  11.69\%        & 2.27\%        & \textbf{16.26\%}       & 5.52\%        & 1.98\%        & 1.63\%        \\
18  & \textbf{19.39\%}         & 11.66\%       & \textbf{12.69\%}         & \textbf{14.84\%}        & 6.97\%  & \textbf{8.53\%}        & \textbf{9.68\%}        & \textbf{4.37\%}        \\ \hline
\end{tabular}

}

\end{table*}
\vspace{-0.05in}
\section{Performance Evaluation}
\label{sec:evaluation}
We evaluate the performance of CEC using real video request traces from two large video streaming platforms. They are from two major companies providing Internet video-on-demand services in China. Videos in the datasets range from movies, TV series, to sports programs and news.
\vspace{-0.05in}
\subsection{Experiment Setup}
\label{sec:setup}
 Dataset A~\cite{li2018data} includes requests generated by users from different provinces in China. Each request record consists of request time, user IP address, content name, valid watching time and video length. The video requests in dataset B~\cite{wang2020intelligent} are from a major city in China. The request record is similar to the first one. Instead of IP address, it identifies each request with user ID and longitude/latitude information.

To emulate different levels of edge caching hierarchy, for dataset A, we group requests by IP prefix. All requests from IP addresses sharing the same prefix at a particular length belong to one edge subnet. The longer the prefix length, the smaller the edge subnet. We use three different prefix lengths: $\{8, 12, 18\}$. For each prefix length, there are many subnets, from which we pick three most active ones. 
For dataset B, we group requests based on geographic information at three levels: \{the whole city, two halves of the city, four quarters of the city\}. For dataset A, there are about \{150k, 100k, 30k\} requests per day for the three levels respectively, while for dataset B, there are  about \{100k, 30k, 20k\} requests per day. In our experiments, for each sub-dataset, we test with three different cache sizes: \{100, 300, 500\} so as to evaluate performance thoroughly. 
For each combination of (\texttt{cache size}, \texttt{subnet level}, \texttt{content provider}), we train a separate DRL model. Each DRL agent is trained using requests from one subnet at that level, and tested on the other subnets at the same level. For the ``whole city" level in dataset B, since there is only one network, we use $70\%$ requests for training, the remaining for testing. We use Python 3.7.6 and PyTorch 1.4.0 to develop the training and testing experiments. The training duration is about half day on a desktop computer with Intel(R) Core(TM) i5-9400 CPU @ 2.90GHz.

Base policies in our ensemble include LFU-$\Delta$ and LRU-$n$ with multiple sliding-window sizes, as well as our proposed LSTM-Int. Since the quality of the policies in the ensemble directly affects the quality of CEC, one may pre-tune parameters of base policies based on their average hit ratios on a small portion of traces before training. In our experiments, we pre-select eight policies, \texttt{LFU-Inf, LFU-500, LFU-5000, LFU-50000, LRU-2, LRU-4, LRU-8} and \texttt{LSTM-Int}, whose average hit ratios on a small portion of the training trace are among the best, to make up the ensemble. {In both training and testing phases, CEC is called once every $100$ requests and the runtime per call is $4.12$ ms on average and the selected policy controls primary cache for the next $100$ requests, which indicates low computation overheads.} We will show next that no single policy dominates the other policies, however, CEC can combine all the pre-selected constituent policies together to boost the hit ratio evidently.

\subsection{Experiment Results}
%


Fig.~\ref{fig:hr_cachesize} shows the average hit ratios of CEC and base policies in all kinds of scenarios of dataset A. Table~\ref{table:relative_increase} is the corresponding relative hit ratio improvement of CEC compared over all the constituent policies. From both Fig.~\ref{fig:hr_cachesize} and Table~\ref{table:relative_increase}, we see CEC has the largest improvement on the smallest edge subnet of prefix length $18$. {\it This echoes our initial motivation that the traditional caching algorithms are inadequate for edge cache boxes serving small user populations.} Note that, the improvement over our proposed LSTM-Int is not as large as over the other policies. This is because LSTM-Int is already a combined policy between LFU and LSTM. The fact that CEC can further improve its performance demonstrates the effectiveness of DRL-based ensemble learning. 


\begin{figure*}[htbp]
\centering{
 \begin{subfigure}[b]{.32\linewidth}
            \includegraphics[width=\linewidth]{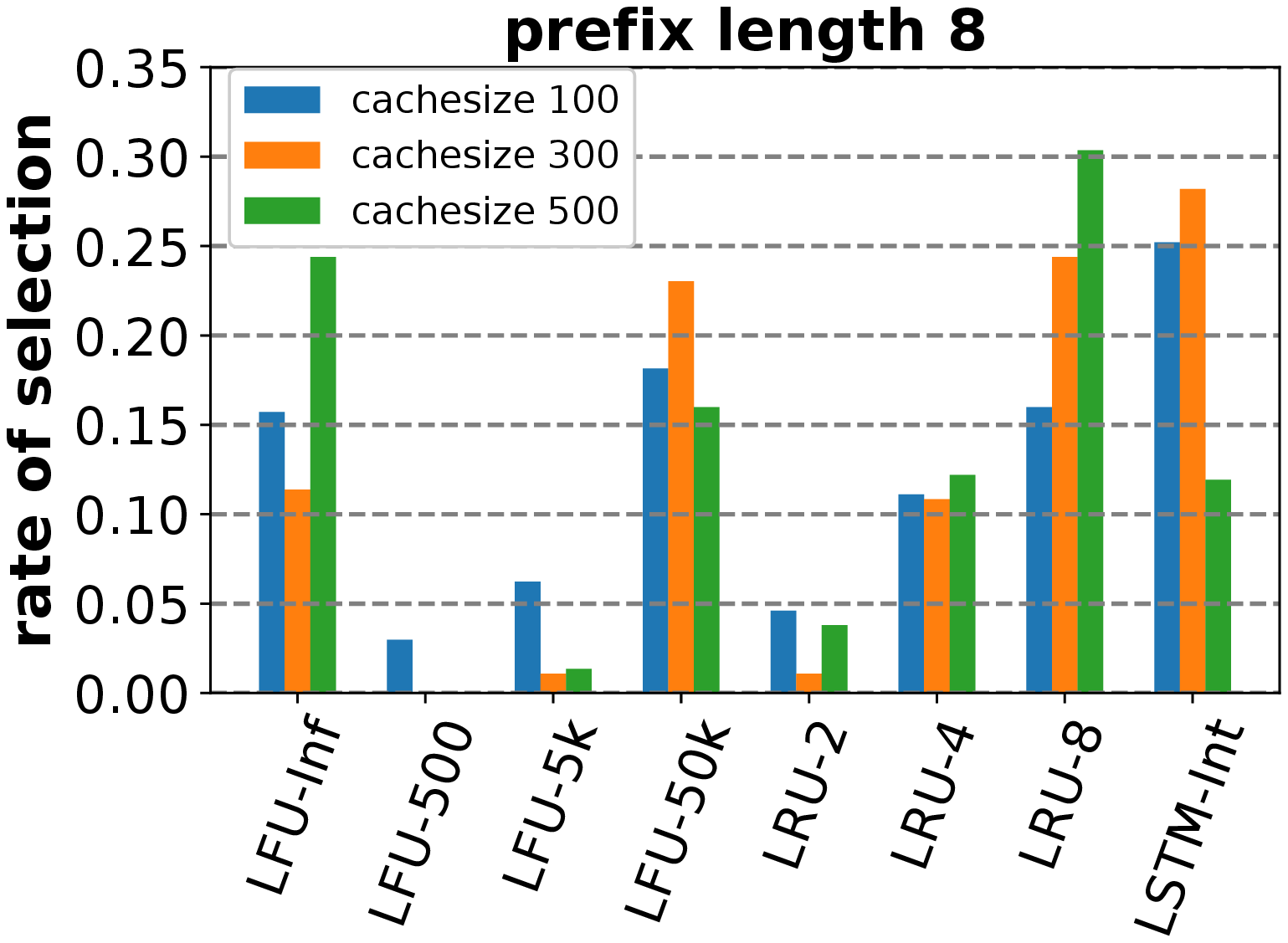}
    \end{subfigure}
    \begin{subfigure}[b]{.32\linewidth}
            \includegraphics[width=\linewidth]{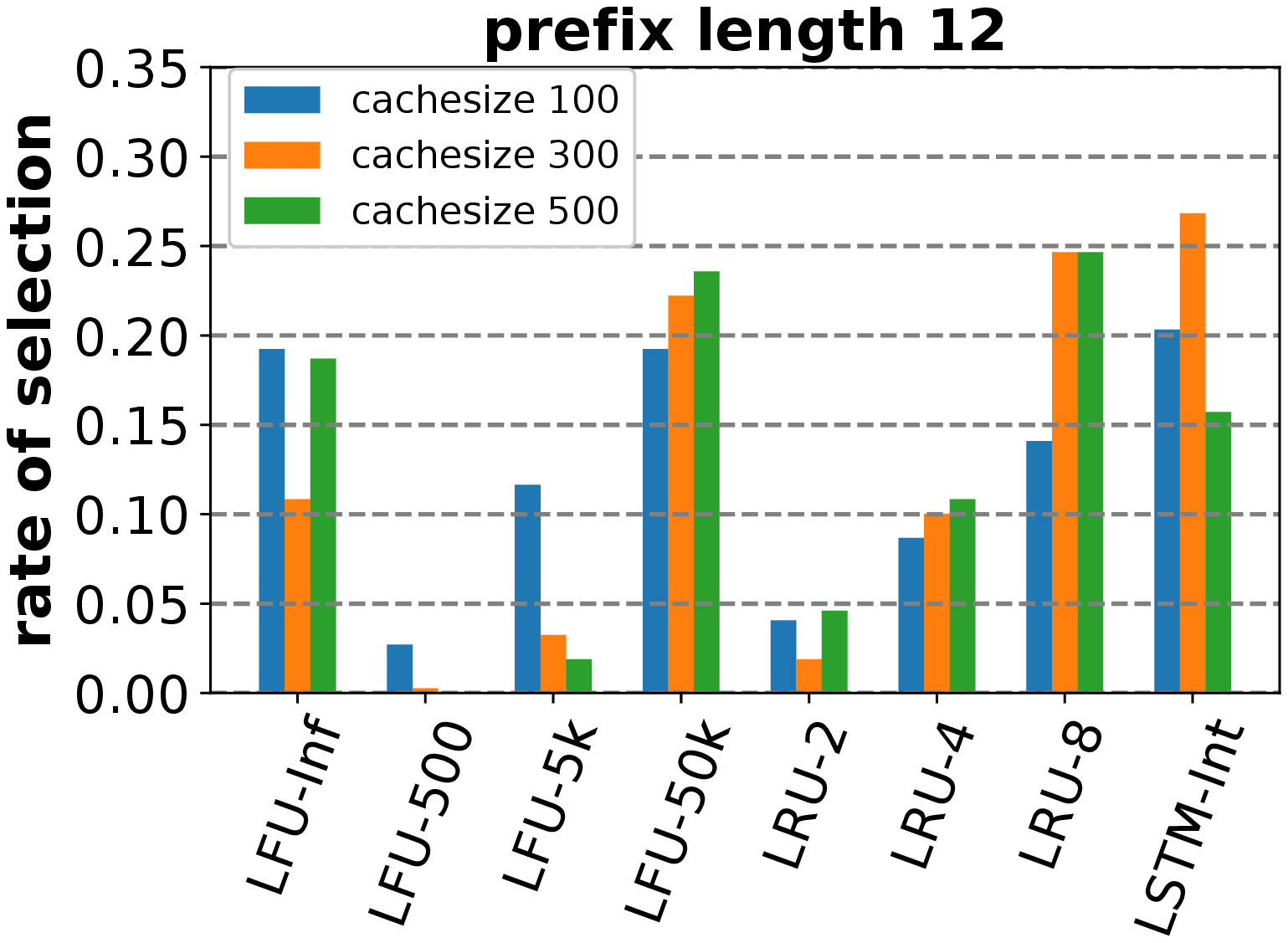}
    \end{subfigure}
    \begin{subfigure}[b]{.32\linewidth}
            \includegraphics[width=\linewidth]{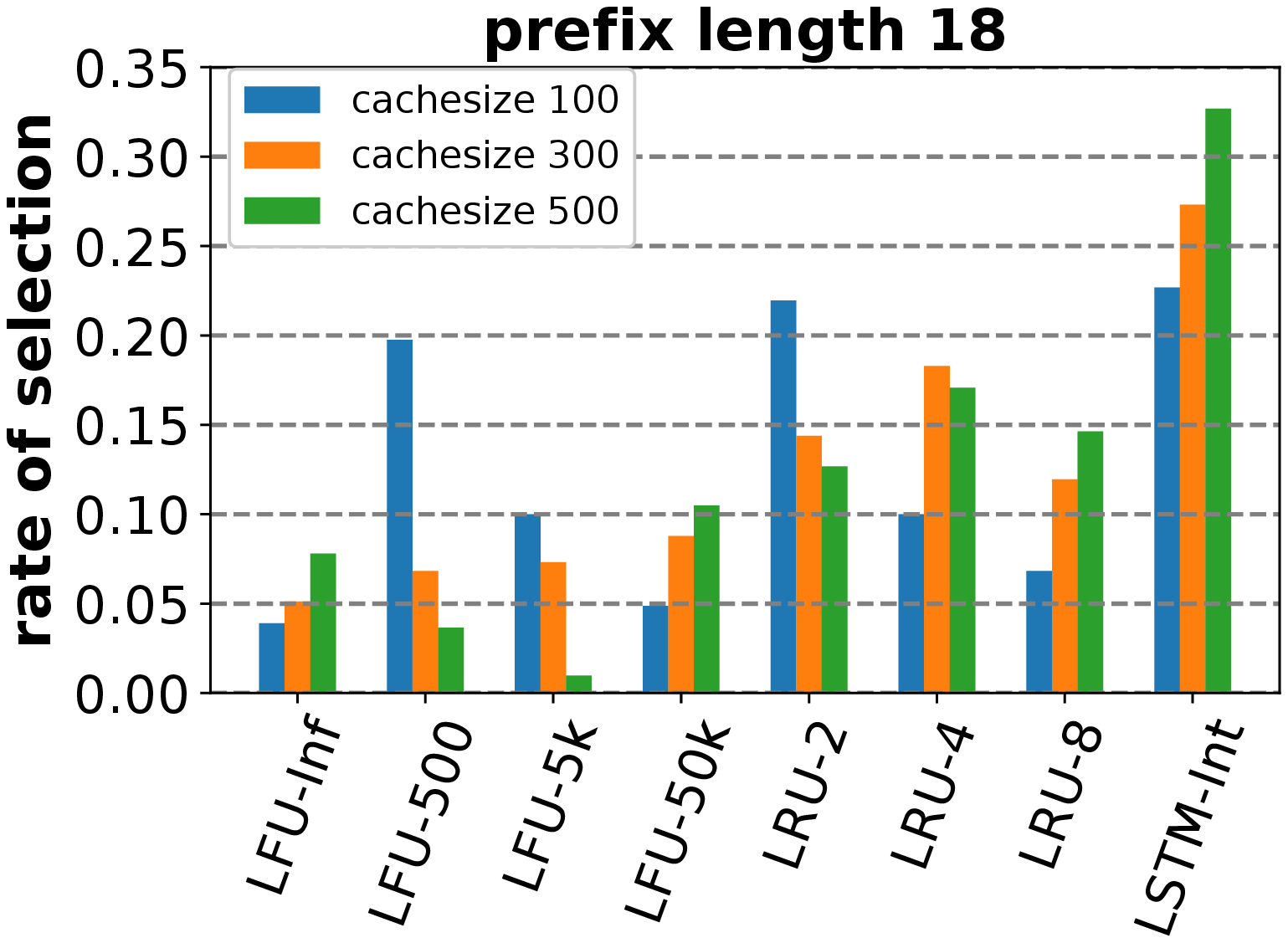}
    \end{subfigure}
}
\caption{\textbf{Rate of Selection} of all policies on dataset A at different prefix lengths}
\label{fig:roe}
\end{figure*}

Fig.~\ref{fig:hr_cachesize} only plots the average hit ratio over the scenarios. To illustrate how CEC adapts to the diurnal patterns in content popularity evolution, we zoom in and create several hundreds of mini-scenarios for different time periods during daytime and nighttime. The duration of each mini-scenario varies from 2 hours to a day. Fig.~\ref{fig:roe} shows how often CEC selects each constituent policy under each cache size and prefix length. y-axis is the fraction of selection of each policy. When observing Fig.~\ref{fig:hr_cachesize} and Fig.~\ref{fig:roe} together, we can see that for a caching scenario, a constituent policy with a higher average hit ratio is more likely to be selected by CEC. And each policy has very different relative performance in the ensemble, and consequently different rates of being selected by CEC, under different scenarios. LSTM-Int performs very well in most cases except when the cache size is 500 at the smallest subnet level. LFU-500 and LRU-2 only play a role when both the subnet size and the  cache size are rather small. LRU-8 performs pretty well when both the subnet and cache size are very large. LFU-50000 is frequently selected, except for the smallest subnet. {\it Due  to the dynamic trends of content popularity, one cannot count on a single policy to deal with all kinds of scenarios. CEC can serve as a general framework to combine different ensembles of base policies pre-selected/customized for different caching scenarios.} 

\begin{table*}[htb]%
\centering
\caption{Relative Hit Ratio Improvement of CEC over Base Policies on Different Scenarios of Dataset B}
  \label{table:iqiyi_result}
{
\begin{tabular}{||c||c|c|c|c|c|c|c|c||}
\hline

\diagbox{}{Baselines}  & & & & & & & & \\
  

 Area  & \bfseries LFU-Inf & \bfseries LFU-500 & \bfseries LFU-5000    & \bfseries LFU-50000  &  \bfseries LRU-2     &  \bfseries LRU-4   & \bfseries LRU-8         & \bfseries LSTM-Int       \\
\hline
whole  & 5.57\%    & \textbf{19.73\%}       & 3.53\%         & 2.41\%   & \textbf{7.56\%}        & 1.71\%        & 1.18\%        & 2.30\%        \\
half  & 14.75\%         &10.27\%  &  4.64\%        & 8.32\%        & 5.31\%       & 2.34\%        & 2.92\%        & 3.71\%        \\
quarter  & \textbf{17.57\%}         & 8.48\%       & \textbf{6.55\%}         & \textbf{11.43\%}        & 4.92\%  & \textbf{3.53\%}        & 4.33\%        & \textbf{5.86\%}        \\ \hline
\end{tabular}

}
\end{table*}


\begin{figure}[htbp]
\centerline{\includegraphics[width=\linewidth]{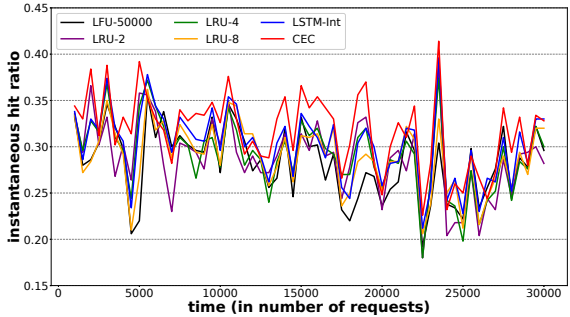}}
\caption{\textbf{Instantaneous Hit Ratios} of CEC and Top-5 base policies on one testing trace}
\label{fig:pHR_time}
\end{figure}



%
In Fig.~\ref{fig:pHR_time}, for a trace from a subnet of prefix $18$ and cache size $500$, we zoom in and plot the temporal evolutions of the instantaneous hit ratios of every $500$ requests for CEC and top-5 base policies for a total of $30,000$ requests. The relative performance of base policies oscillates overtime, without a dominating single policy. CEC always picks the most suitable policy and maintains the highest hit ratio almost all the time.

\vspace{-0.2in}
\begin{figure}[htbp]
\centerline{\includegraphics[width=\linewidth]{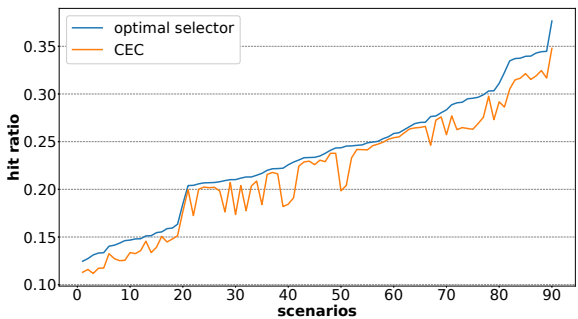}}
\caption{Hit Ratios of Optimal Selector and CEC}
\label{fig:upperbound}
\end{figure}

To assess the quality of CEC policy selection, we compare it with an unrealistic optimal policy selector that has access to the ``oracle" of future request arrival times. For each new request, the optimal selector calculates the FIF scores according to (\ref{eq:fif_score}) of all the eviction candidates chosen by all the base policies, and evicts the one with the lowest FIF score. 
Fig.~\ref{fig:upperbound} shows the hit ratios across almost one hundred mini-scenarios. The x-axis represents the mini-scenario index, sorted in the increasing order of the hit ratios of FIF policy selector. 
For some easy scenarios, e.g. the popularity of contents is relatively stable, CEC can closely match FIF. However, big gaps are also observed for some more complicated scenarios. There are several reasons: 1) FIF has access to the oracle of future request arrivals; 2) FIF can change policy for each request,  whereas CEC only changes policy once every 100 requests; 3) DRL agent might need additional informative context information to make better selection. We will improve CEC to close the gap in future. 
  
Due to the space limit, for dataset B, we only show the relative improvement of CEC over base policies for different scales of geographic areas: the whole city, a half of the city, and a quarter of the city in Table~\ref{table:iqiyi_result}. The overall trends are quite similar to dataset A.  CEC achieves the largest improvement on the smallest area scale, which further demonstrates the advantages of CEC for edge caching.
\vspace{-0.05in}
\section{Conclusion}
In this paper, we studied edge caching policies using ensemble learning. We first showed through analysis and experiments that the classical LFU and LRU policies  can be enhanced with more flexible time series analysis, in particular LSTM based future popularity prediction to  adapt better to dynamic content popularity and heterogeneous caching scenarios. We then developed a novel Cocktail Edge Caching framework that strategically combines base caching policies using DRL to achieve superior and more robust performance than any single policy. In future, we will improve CEC by incorporating more informative context features and adopting the constantly evolving new DRL models.    \newpage   







\bibliographystyle{IEEEtran}
\bibliography{ref,mpip_p2,SN-Recom,360_New,nets14}

\begin{thebibliography}{10}
\providecommand{\url}[1]{#1}
\csname url@samestyle\endcsname
\providecommand{\newblock}{\relax}
\providecommand{\bibinfo}[2]{#2}
\providecommand{\BIBentrySTDinterwordspacing}{\spaceskip=0pt\relax}
\providecommand{\BIBentryALTinterwordstretchfactor}{4}
\providecommand{\BIBentryALTinterwordspacing}{\spaceskip=\fontdimen2\font plus
\BIBentryALTinterwordstretchfactor\fontdimen3\font minus
  \fontdimen4\font\relax}
\providecommand{\BIBforeignlanguage}[2]{{%
\expandafter\ifx\csname l@#1\endcsname\relax
\typeout{** WARNING: IEEEtran.bst: No hyphenation pattern has been}%
\typeout{** loaded for the language `#1'. Using the pattern for}%
\typeout{** the default language instead.}%
\else
\language=\csname l@#1\endcsname
\fi
#2}}
\providecommand{\BIBdecl}{\relax}
\BIBdecl

\bibitem{Huawei_Report}
Huawei, ``{Whitepaper on the VR-Oriented Bearer Network Requirement(2016)},''
  Huawei Technology, Tech. Rep., 2016, available at
  \url{http://www-file.huawei.com/~/media/CORPORATE/PDF/white%20paper/whitepaper-on-the-vr-oriented-bearer-network-requirement-en.pdf}.

\bibitem{In-network}
I.~Psaras, W.~K. Chai, and G.~Pavlou, ``Probabilistic in-network caching for
  information-centric networks,'' in \emph{Proceedings of the Second Edition of
  the ICN Workshop on Information-centric Networking}, ser. ICN '12, 2012, pp.
  55--60.

\bibitem{jacobson2009networking}
V.~Jacobson, D.~K. Smetters, J.~D. Thornton, M.~F. Plass, N.~H. Briggs, and
  R.~L. Braynard, ``Networking named content,'' in \emph{Proceedings of the 5th
  international conference on Emerging networking experiments and
  technologies}.\hskip 1em plus 0.5em minus 0.4em\relax ACM, 2009, pp. 1--12.

\bibitem{FemtoCaching}
N.~Golrezaei, K.~Shanmugam, A.~G. Dimakis, A.~F. Molisch, and G.~Caire,
  ``Femtocaching: Wireless video content delivery through distributed caching
  helpers,'' in \emph{Proc. IEEE INFOCOM}.\hskip 1em plus 0.5em minus
  0.4em\relax IEEE, 2012.

\bibitem{sun2020flocking}
L.~Sun, Y.~Mao, T.~Zong, Y.~Liu, and Y.~Wang, ``Flocking-based live streaming
  of 360-degree video,'' in \emph{Proceedings of the 11th ACM Multimedia
  Systems Conference}, 2020, pp. 26--37.

\bibitem{Sagi2018ensemble}
O.~Sagi and L.~Rokach, ``Ensemble learning: A survey,'' \emph{WIREs Data Mining
  and Knowledge Discovery}, vol.~8, no.~4, p. e1249, 2018.

\bibitem{NetflixPrize}
\BIBentryALTinterwordspacing
NetFlix. Netflix prize challenge wikipedia. [Online]. Available:
  \url{https://en.wikipedia.org/wiki/Netflix_Prize}
\BIBentrySTDinterwordspacing

\bibitem{WinningNetflix}
R.~Bell, Y.~Koren, and C.~Volinsky, ``The bellkor 2008 solution to the netflix
  prize,'' \emph{AT\&T Research}, 01 2008.

\bibitem{henkel1999attacking}
J.~Henkel, ``Attacking aids with acocktail'therapy.'' \emph{FDA consumer},
  vol.~33, no.~4, pp. 12--17, 1999.

\bibitem{cisco2018cisco}
V.~Cisco, ``Cisco visual networking index: Forecast and trends, 2017--2022,''
  \emph{White Paper}, 2018.

\bibitem{Fofack2012}
N.~C. {Fofack}, P.~{Nain}, G.~{Neglia}, and D.~{Towsley}, ``Analysis of
  ttl-based cache networks,'' in \emph{6th International ICST Conference on
  Performance Evaluation Methodologies and Tools}, 2012, pp. 1--10.

\bibitem{Melazzi2014}
N.~B. {Melazzi}, G.~{Bianchi}, A.~{Caponi}, and A.~{Detti}, ``A general,
  tractable and accurate model for a cascade of lru caches,'' \emph{IEEE
  Communications Letters}, vol.~18, no.~5, pp. 877--880, 2014.

\bibitem{ChiaTaiChan2000}
{Chia-Tai Chan}, {Shuo-Cheng Hu}, {Pi-Chung Wang}, and {Yaw-Chung Chen}, ``A
  fifo-based buffer management approach for the atm gfr services,'' \emph{IEEE
  Communications Letters}, vol.~4, no.~6, pp. 205--207, 2000.

\bibitem{cherkasova1998improving}
L.~Cherkasova, \emph{Improving WWW proxies performance with
  greedy-dual-size-frequency caching policy}.\hskip 1em plus 0.5em minus
  0.4em\relax Hewlett-Packard Laboratories, 1998.

\bibitem{IRM}
E.~G. Coffman, Jr. and P.~J. Denning, \emph{Operating Systems Theory}.\hskip
  1em plus 0.5em minus 0.4em\relax Prentice Hall Professional Technical
  Reference, 1973.

\bibitem{IRM1}
P.~R. Jelenkovic and A.~Radovanovic, ``The persistent-access-caching
  algorithm,'' \emph{Random Struct. Algorithms}, vol.~33, no.~2, pp. 219--251,
  2008.

\bibitem{IRM2}
A.~Dan and D.~Towsley, ``An approximate analysis of the lru and fifo buffer
  replacement schemes,'' in \emph{Proceedings of the 1990 ACM SIGMETRICS
  Conference on Measurement and Modeling of Computer Systems}, ser. SIGMETRICS
  '90, 1990.

\bibitem{li2016popularity}
S.~Li, J.~Xu, M.~Van Der~Schaar, and W.~Li, ``Popularity-driven content
  caching,'' in \emph{IEEE INFOCOM 2016-The 35th Annual IEEE International
  Conference on Computer Communications}.\hskip 1em plus 0.5em minus
  0.4em\relax IEEE, 2016, pp. 1--9.

\bibitem{song2017learning}
J.~Song, M.~Sheng, T.~Q. Quek, C.~Xu, and X.~Wang, ``Learning-based content
  caching and sharing for wireless networks,'' \emph{IEEE Transactions on
  Communications}, vol.~65, no.~10, pp. 4309--4324, 2017.

\bibitem{Ma2017}
G.~{Ma}, Z.~{Wang}, M.~{Zhang}, J.~{Ye}, M.~{Chen}, and W.~{Zhu},
  ``Understanding performance of edge content caching for mobile video
  streaming,'' \emph{IEEE Journal on Selected Areas in Communications},
  vol.~35, no.~5, pp. 1076--1089, 2017.

\bibitem{Zhang2019}
C.~{Zhang}, H.~{Pang}, J.~{Liu}, S.~{Tang}, R.~{Zhang}, D.~{Wang}, and
  L.~{Sun}, ``Toward edge-assisted video content intelligent caching with long
  short-term memory learning,'' \emph{IEEE Access}, vol.~7, pp.
  152\,832--152\,846, 2019.

\bibitem{li2018data}
G.~Li, Q.~Shen, Y.~Liu, H.~Cao, Z.~Han, F.~Li, and J.~Li, ``Data-driven
  approaches to edge caching,'' in \emph{Proceedings of the 2018 Workshop on
  Networking for Emerging Applications and Technologies}, 2018, pp. 8--14.

\bibitem{poularakis2016}
K.~{Poularakis}, G.~{Iosifidis}, A.~{Argyriou}, I.~{Koutsopoulos}, and
  L.~{Tassiulas}, ``Caching and operator cooperation policies for layered video
  content delivery,'' in \emph{IEEE INFOCOM 2016 - The 35th Annual IEEE
  International Conference on Computer Communications}, 2016, pp. 1--9.

\bibitem{Wang2017}
S.~{Wang}, X.~{Zhang}, Y.~{Zhang}, L.~{Wang}, J.~{Yang}, and W.~{Wang}, ``A
  survey on mobile edge networks: Convergence of computing, caching and
  communications,'' \emph{IEEE Access}, vol.~5, pp. 6757--6779, 2017.

\bibitem{Andre2017edgecache}
\BIBentryALTinterwordspacing
A.~S. Gomes, B.~Sousa, D.~Palma, V.~Fonseca, Z.~Zhao, E.~Monteiro, T.~Braun,
  P.~Simoes, and L.~Cordeiro, ``Edge caching with mobility prediction in
  virtualized lte mobile networks,'' \emph{Future Generation Computer Systems},
  vol.~70, pp. 148 -- 162, 2017. [Online]. Available:
  \url{http://www.sciencedirect.com/science/article/pii/S0167739X16302072}
\BIBentrySTDinterwordspacing

\bibitem{somuyiwa2018reinforcement}
S.~O. Somuyiwa, A.~Gy{\"o}rgy, and D.~G{\"u}nd{\"u}z, ``A
  reinforcement-learning approach to proactive caching in wireless networks,''
  \emph{IEEE Journal on Selected Areas in Communications}, vol.~36, no.~6, pp.
  1331--1344, 2018.

\bibitem{zhong2018deep}
C.~Zhong, M.~C. Gursoy, and S.~Velipasalar, ``A deep reinforcement
  learning-based framework for content caching,'' in \emph{2018 52nd Annual
  Conference on Information Sciences and Systems (CISS)}.\hskip 1em plus 0.5em
  minus 0.4em\relax IEEE, 2018, pp. 1--6.

\bibitem{Kirilin2020}
V.~{Kirilin}, A.~{Sundarrajan}, S.~{Gorinsky}, and R.~K. {Sitaraman},
  ``Rl-cache: Learning-based cache admission for content delivery,'' \emph{IEEE
  Journal on Selected Areas in Communications}, pp. 1--1, 2020.

\bibitem{fan2020pa}
Q.~Fan, J.~Li, X.~Li, Q.~He, and S.~Fu, ``Pa-cache: Learning-based
  popularity-aware content caching in edge networks,'' 02 2020.

\bibitem{sadeghi2019reinforcement}
A.~Sadeghi, F.~Sheikholeslami, A.~G. Marques, and G.~B. Giannakis,
  ``Reinforcement learning for adaptive caching with dynamic storage pricing,''
  \emph{IEEE Journal on Selected Areas in Communications}, vol.~37, no.~10, pp.
  2267--2281, 2019.

\bibitem{sengupta2014learning}
A.~Sengupta, S.~Amuru, R.~Tandon, R.~M. Buehrer, and T.~C. Clancy, ``Learning
  distributed caching strategies in small cell networks,'' in \emph{2014 11th
  International Symposium on Wireless Communications Systems (ISWCS)}.\hskip
  1em plus 0.5em minus 0.4em\relax IEEE, 2014, pp. 917--921.

\bibitem{jiang2019multi}
W.~Jiang, G.~Feng, S.~Qin, T.~S.~P. Yum, and G.~Cao, ``Multi-agent
  reinforcement learning for efficient content caching in mobile d2d
  networks,'' \emph{IEEE Transactions on Wireless Communications}, vol.~18,
  no.~3, pp. 1610--1622, 2019.

\bibitem{sadeghi2017optimal}
A.~Sadeghi, F.~Sheikholeslami, and G.~B. Giannakis, ``Optimal and scalable
  caching for 5g using reinforcement learning of space-time popularities,''
  \emph{IEEE Journal of Selected Topics in Signal Processing}, vol.~12, no.~1,
  pp. 180--190, 2017.

\bibitem{Zhu2018}
H.~{Zhu}, Y.~{Cao}, W.~{Wang}, T.~{Jiang}, and S.~{Jin}, ``Deep reinforcement
  learning for mobile edge caching: Review, new features, and open issues,''
  \emph{IEEE Network}, vol.~32, no.~6, pp. 50--57, 2018.

\bibitem{Wu2019}
P.~{Wu}, J.~{Li}, L.~{Shi}, M.~{Ding}, K.~{Cai}, and F.~{Yang}, ``Dynamic
  content update for wireless edge caching via deep reinforcement learning,''
  \emph{IEEE Communications Letters}, vol.~23, no.~10, pp. 1773--1777, 2019.

\bibitem{wang2020intelligent}
F.~Wang, F.~Wang, J.~Liu, R.~Shea, and L.~Sun, ``Intelligent video caching at
  network edge: A multi-agent deep reinforcement learning approach,'' in
  \emph{IEEE INFOCOM 2020-IEEE Conference on Computer Communications}.\hskip
  1em plus 0.5em minus 0.4em\relax IEEE, 2020, pp. 2499--2508.

\bibitem{traverso2015unravelling}
S.~Traverso, M.~Ahmed, M.~Garetto, P.~Giaccone, E.~Leonardi, and S.~Niccolini,
  ``Unravelling the impact of temporal and geographical locality in content
  caching systems,'' \emph{IEEE Transactions on Multimedia}, vol.~17, no.~10,
  pp. 1839--1854, 2015.

\bibitem{traverso2013temporal}
------, ``Temporal locality in today's content caching: why it matters and how
  to model it,'' \emph{ACM SIGCOMM Computer Communication Review}, vol.~43,
  no.~5, pp. 5--12, 2013.

\bibitem{Belady1966}
L.~{Belady}, ``A study of replacement algorithms for a virtual-storage
  computer,'' \emph{IBM Systems journal}, vol.~5, no.~2, pp. 78--101, 1966.

\bibitem{brown1992introduction}
R.~G. Brown and P.~Y.~C. Hwang, \emph{{Introduction to random signals and
  applied kalman filtering: with MATLAB exercises and solutions; 3rd
  ed.}}\hskip 1em plus 0.5em minus 0.4em\relax New York, NY: Wiley, 1997.

\bibitem{haykin2008adaptive}
S.~Haykin, \emph{Adaptive Filter Theory (3rd Ed.)}.\hskip 1em plus 0.5em minus
  0.4em\relax Upper Saddle River, NJ, USA: Prentice-Hall, Inc., 1996.

\bibitem{gers1999learning}
F.~A. Gers, J.~A. Schmidhuber, and F.~A. Cummins, ``Learning to forget:
  Continual prediction with lstm,'' \emph{Neural Comput.}, vol.~12, no.~10,
  Oct. 2000.

\bibitem{li2018accurate}
J.~Li, S.~Shakkottai, J.~C. Lui, and V.~Subramanian, ``Accurate learning or
  fast mixing? dynamic adaptability of caching algorithms,'' \emph{IEEE Journal
  on Selected Areas in Communications}, vol.~36, no.~6, pp. 1314--1330, 2018.

\bibitem{megiddo2003arc}
N.~Megiddo and D.~S. Modha, ``Arc: A self-tuning, low overhead replacement
  cache.'' in \emph{Fast}, vol.~3, no. 2003, 2003, pp. 115--130.

\bibitem{van2016deep}
H.~Van~Hasselt, A.~Guez, and D.~Silver, ``Deep reinforcement learning with
  double q-learning,'' in \emph{Thirtieth AAAI conference on artificial
  intelligence}, 2016.

\end{thebibliography}

\end{document}